\input{aipcheck.tex}
\documentclass{aipproc}
\layoutstyle{6x9}

\begin{document}
\def\dsl{\,\raise.15ex\hbox{/}\mkern-13.5mu D} 
\def\xslash#1{{\rlap{$#1$}/}}
\def\openone{\leavevmode\hbox{\small1\kern-4.2pt\normalsize1}}
\def\ssqr#1#2{{\vbox{\hrule height #2pt
      \hbox{\vrule width #2pt height#1pt \kern#1pt\vrule width #2pt}
      \hrule height #2pt}\kern- #2pt}}
\def\sqr{\mathchoice\ssqr8{.4}\ssqr8{.4}\ssqr{5}{.3}\ssqr{4}{.3}}
\def\onedot{\makebox(0,0){$\scriptstyle 1$}}
\def\twodot{\makebox(0,0){$\scriptstyle 2$}}
\def\threedot{\makebox(0,0){$\scriptstyle 3$}}
\def\fourdot{\makebox(0,0){$\scriptstyle 4$}}
\def\Tr{{\rm Tr}}
\def\onebox{{\vbox{\hbox{$\sqr\thinspace$}}}}
\def\twobox{{\vbox{\hbox{$\sqr\sqr\thinspace$}}}}
\def\threebox{{\vbox{\hbox{$\sqr\sqr\sqr\thinspace$}}}}
\def\nbox{\hbox{$\sqr\sqr\sqr\sqr\raise2.7pt\hbox{$\,\cdot\cdot\cdot
\cdot\cdot\,$}\sqr\sqr\sqr\thinspace$}}
\def\nboxE{\vbox{\hbox{$\sqr\sqr\sqr\raise2.7pt\hbox{$\,\cdot\cdot\cdot
\cdot\cdot\,$}\sqr\sqr\sqr\sqr$}\nointerlineskip 
\kern-.2pt\hbox{$\sqr\sqr\sqr\raise2.7pt\hbox{$\,\cdot\cdot\cdot
\cdot\cdot\,$}\sqr$}}}
\def\nboxF{\vbox{\hbox{$\sqr\sqr\sqr\sqr\raise2.7pt\hbox{$\,\cdot\cdot\cdot
\cdot\cdot\,$}\sqr\sqr$}\nointerlineskip 
\kern-.2pt\hbox{$\sqr\sqr\sqr\sqr\raise2.7pt\hbox{$\,\cdot\cdot\cdot
\cdot\cdot\,$}\sqr$}}}

\title
[QCD Baryons in the $1/N_c$ Expansion]{QCD Baryons in the $1/N_c$ Expansion}

\author{Elizabeth Jenkins}{
  address={Department of Physics, 9500 Gilman Drive, 
  University of California at San Diego, La Jolla, CA 92093-0319},
}

\copyrightyear  {2001}

\begin{abstract}
The $1/N_c$ expansion provides a theoretical method for analyzing the 
spin-flavor symmetry properties of baryons in QCD that is quantitative, 
systematic and predictive.  An exact spin-flavor symmetry 
exists for large-$N_c$ baryons, whereas for QCD baryons, the spin-flavor 
symmetry is approximate and is broken by corrections proportional to the 
symmetry-breaking parameter $1/N_c = 1/3$.  The $1/N_c$ expansion
predicts a hierarchy of spin and flavor symmetry relations for QCD baryons
that is observed in nature.  It provides a quantitative understanding of
why some $SU(3)$ flavor symmetry relations in the baryon sector, such as the 
Gell-Mann--Okubo mass formula, are satisfied to a greater precision than
expected from flavor symmetry-breaking suppression factors alone. 
\end{abstract}

\date{\today}

\maketitle

\section{Introduction}

It has been rigorously proven that spin-flavor
symmetry is an approximate symmetry of baryons in QCD \cite{dm,j}.  
Spin-flavor symmetry for baryons is formally an exact symmetry in the 
t'Hooft large-$N_c$ limit~\cite{thooft}.  For finite $N_c$, the 
spin-flavor symmetry of baryons is only approximate and is broken
explicitly by corrections suppressed by powers of $1/N_c$.
The breaking of the large-$N_c$ baryon
spin-flavor symmetry for QCD baryons is order $1/N_c = 1/3$, which is 
comparable to the order $30\%$ breaking of Gell-Mann $SU(3)$ flavor symmetry.  
Thus, spin-flavor symmetry is as good an approximate symmetry for QCD baryons 
as $SU(3)$ flavor symmetry.  

Spin-flavor symmetry for QCD baryons has a long history; like $SU(3)$ flavor symmetry,
spin-flavor symmetry predates the formulation of QCD.  Although
spin-flavor symmetry was phenomenologically
successful early on \cite{oldsu6}, the physical basis for spin-flavor symmetry was not understood, 
even after the microscopic theory of the strong interactions was known.
What has been possible in recent years is to justify spin-flavor symmetry as a {\it bona fide}
symmetry of baryons in QCD, and to classify the explicit breakings of baryon spin-flavor
symmetry to all orders in the $1/N_c$ expansion in a systematic and quantitative
manner.  It has been shown that the quark-gluon dynamics of large-$N_c$ QCD gives rise to 
a spin-flavor symmetry for baryons.  For finite $N_c$, the symmetry is only approximate; there are 
subleading $1/N_c$ corrections which explicitly break the symmetry.  This new insight 
has led to the formulation of the baryon $1/N_c$
expansion as an expansion in operators with definite transformation
properties under baryon spin and flavor symmetry.  Each baryon operator in the $1/N_c$ expansion
occurs at a known order in $1/N_c$.  Each operator is multiplied by an unknown coefficient
which is a reduced matrix element that is not determined by baryon spin-flavor symmetry.  
Calculating these reduced baryon matrix elements is tantamount
to solving QCD in the baryon sector.

All of the new results obtained for baryons in the $1/N_c$
expansion can be characterized as symmetry relations involving spin and flavor.  The
spin-flavor structure of the baryon $1/N_c$ expansion yields model-independent results 
which are valid for QCD.  It has been known for some time that the spin-flavor group theory for
baryons in large-$N_c$ QCD and in the large-$N_c$ quark and Skyrme models is the 
same\cite{manohar84}.
It is now known that a stronger statement applies:   
the spin-flavor structure of the baryon $1/N_c$ expansion is the same in QCD
as in the quark model and the Skyrme model at each order in the $1/N_c$ expansion.
In other words, 
the spin-flavor structure of the baryon $1/N_c$ expansion is the same in QCD
as in the quark model and Skyrme models.  QCD and these models, however, will differ in their
predictions for the reduced matrix elements of the baryon $1/N_c$ expansion.      
It is well known that the quark and
Skyrme models are not very successful at predicting these reduced matrix 
elements, 
which
leads to the conclusion that most, if not all, of the successful predictions of the 
non-relativistic quark and Skyrme models are
actually model-independent group-theoretic predictions, and therefore should not be regarded
as evidence for the validity of these models {\it per se}.

The spin-flavor operator analysis of the
baryon $1/N_c$ expansion has resulted in significant progress in understanding of the
spin-flavor structure of baryons.  It explains the extraordinary accuracy of many venerable
spin-flavor and flavor symmetry relations for baryons 
as being due to the presence of $1/N_c$ suppression 
factors in addition to the usual flavor symmetry breaking suppression factors.  It is
remarkable that, after three decades, a  
quantitative understanding of spin-flavor symmetry for baryons has been achieved. 

The outline of these lectures is as follows.  First, a brief
summary is given of the $1/N_c$ expansion of large-$N_c$ QCD.  The $1/N_c$ power counting
of the quark-gluon dynamics of large-$N_c$ QCD is described for the confined quark-gluon bound states:
mesons and baryons.  Second, it is shown that large-$N_c$ baryons satisfy a
contracted spin-flavor symmetry in the $N_c \rightarrow \infty$ limit.
Baryon states transform as irreducible representations of the
spin-flavor algebra, and operators acting on a baryon spin-flavor multiplet
transform as irreducible tensor operators of the algebra. 

The formalism of the baryon $1/N_c$ expansion is presented in the next section.
Tensor operators acting on a baryon spin-flavor multiplet have an
expansion in terms of operator products of the baryon spin-flavor generators.  The order
in $1/N_c$ of each operator product in the $1/N_c$ expansion
is known.  Not all operator
products of the baryon spin-flavor generators are linearly independent, so it is necessary
to eliminate redundant operator products using operator identities.  This operator reduction
is possible in terms of the operator identities for $2$-body operator products.  The complete set 
of $2$-body operator product identities for $SU(6)$ spin-flavor symmetry is given, and then
used to construct operator bases for the baryon $1/N_c$ expansion.   

$1/N_c$ operator-product expansions are constructed for a number of baryon tensor operators
in the final section.
It is necessary to incorporate $SU(3)$ flavor symmetry breaking into the baryon $1/N_c$
expansion since $SU(3)$ breaking is comparable to the expansion parameter $1/N_c$. 
The $1/N_c$ expansions for baryon masses, axial vector couplings and magnetic moments
are presented in detail.  A comparison of experimental data with the predictions of the combined
$1/N_c$ and flavor-symmetry breaking expansion is given in these cases.  
The presence of $1/N_c$
suppression factors in the experimental data is clearly evident, and provides a quantitative
understanding of the accuracy of famous symmetry relations, such as the Gell-Mann--Okubo 
formula, Gell-Mann's Equal Spacing Rule and the Coleman-Glashow relation for baryon masses,
as well as many others.

The focus of these lectures is on the application of the $1/N_c$ expansion to QCD baryons. 
The discussion is
meant to be complementary to my review of large-$N_c$ baryons \cite{arnps}, which 
is more comprehensive.  The emphasis here is on
baryons in QCD with $N_F =3$ flavors of light quarks.     
I will try to keep the formalism introduced to a 
minimum throughout, although a fair amount is essential and cannot be avoided.
Of necessity, a number of important topics have not been covered, even briefly.
These topics include nonet symmetry of baryon amplitudes \cite{jchpt}, exact cancellations in
baryon chiral perturbation theory \cite{fhjm}, spin-flavor symmetry of excited baryons
\cite{cgkm}, and
spin-flavor symmetry of heavy quark baryons \cite{jhqet}.  
An extensive list of publications on the spin and flavor properties of baryons in
the $1/N_c$ expansion is given in the references.    

\section{Large-$N_c$ QCD}

A recent review of large-$N_c$ QCD can be found in Ref.~\cite{leshouches}, so the presentation
here will be brief.

Large-$N_c$ QCD is defined as the generalization of $SU(3)$ gauge theory of
quarks and gluons to $SU(N_c)$ gauge theory.  The naive generalization of the QCD Lagrangian
is given by
\begin{equation}\label{lag}
{\cal L} = -\frac 1 2 {\rm Tr}\ G^{\mu \nu} G_{\mu \nu} 
+ \sum_{f=1}^{N_F}\bar q_f \left( i\ \xslash D - m_f \right) q_f,
\end{equation}
where the gauge field strength and the covariant derivative, 
$D^\mu = \partial^\mu + i g A^\mu$, are defined as
in QCD.  For $SU(N_c)$ gauge theory,
the gluons appear in the adjoint represention of $SU(N_c)$ with dimension $(N_c^2 -1)$
while the quarks appear in the fundamental representation $\bf N_c$.  Thus, there are
$O(N_c)$ more gluon degrees of freedom than quark degrees of freedom in large-$N_c$ QCD.  

\begin{figure}\label{doubleline}
\resizebox{.25\textwidth}{!}
{\includegraphics{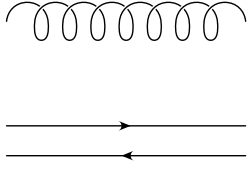}}
\caption{Double line notation for a gluon.}
\end{figure}

The $1/N_c$ power
counting of quark-gluon diagrams is readily obtained by introducing t'Hooft double line
notation for the gluon gauge field:  the adjoint index $A$ on the gauge field $\left(A^\mu
\right)^A$ is replaced by fundamental and anti-fundamental indices $i$ and $j$,
so that the gauge field is written as $\left( A^\mu \right)^i_j$, a substitution which is valid
up to corrections which are subleading in the $1/N_c$ expansion.  In double line notation,
the gluon is effectively replaced by a quark line and an antiquark line, as shown in Fig.~1.
When a quark-gluon Feynman diagram is rewritten in double line notation, determining the power in
$N_c$ of the diagram is equivalent to counting the number of closed quark loops with
unrestricted color summations.  Fig.~2 gives an explicit example of this $1/N_c$ counting.  The
diagram shown in Fig.~2 reduces to three quark
loops, and is therefore proportional to $N_c^3$.  In addition, the diagram is proportional to 
four powers of the quark-gluon coupling constant $g$, since the diagram contains four vertices,
each of which is proportional to $g$.  Thus, the overall diagram is proportional to $g^4 N_c^3
= \left(g^2 N_c \right)^2 N_c$.  A simple analysis of other diagrams leads to the
t'Hooft result that vacuum Feynman diagrams are proportional to
\begin{equation}
\left(g^2 N_c\right)^{\frac 1 2 V_3 + V_4} N_c^{\chi},
\end{equation}
where $V_n$ is the number of $n$-point
vertices in the diagram and $\chi$ is the Euler character of the diagram.  Consequently,
diagrams with arbitrary numbers of $3$- and $4$-point vertices grow with arbitrarily
large powers of $N_c$ unless the limit $N_c \rightarrow \infty$ is taken with
$g^2 N_c$ held fixed.  This limiting procedure, which is necessary to define $SU(N_c)$
gauge theory in the large-$N_c$ limit, is known as the t'Hooft limit.  The constraint $g^2 N_c$
held fixed can be implemented by rescaling the gauge coupling $g \rightarrow g/\sqrt{N_c}$
in the original Lagrangian.  After this rescaling, Feynman diagrams will be proportional to
$N_c^\chi$, where the Euler character $\chi= 2 -2H -L$ can be computed in terms of the number
of handles $H$ and quark loops $L$ of a given diagram.  The t'Hooft limit leads to the
following results:
\begin{itemize}
\item For finite and large $N_c$, planar diagrams with $H=0$ dominate the dynamics.  (All planar
diagrams with a given $L$ are of the same order.)  
\item Diagrams with nonplanar gluon exchange ($H\ne 0$) are suppressed relative to
planar diagrams by one factor of $1/N_c^2$ for each nonplanar gluon.
\item Diagrams with quark loops ($L\ne 0$) are suppressed by one factor of $1/N_c$ for each
quark loop.
\end{itemize}

\begin{figure}\label{threeloop}
\resizebox{.5\textwidth}{!}
{\includegraphics{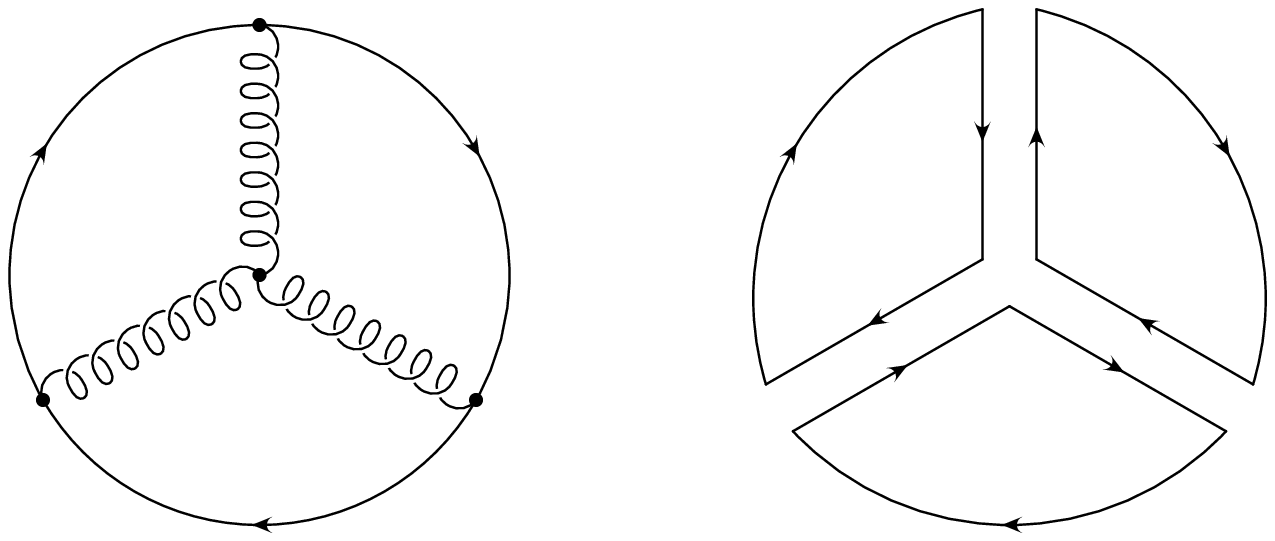}}
\caption{}
\end{figure}

The dynamics of large-$N_c$ QCD is presumed to be confining.  The $\beta$ function of 
large-$N_c$ QCD implies that the rescaled coupling gets large at some scale, let us call it
$\Lambda_{\rm QCD}$.  For $E \le O(\Lambda_{\rm QCD})$, large-$N_c$ QCD is 
strongly coupled and is expected to exhibit confinement.  The confined theory contains colorless bound 
states: mesons, baryons and glueballs.  The $1/N_c$ power counting for large-$N_c$ mesons and
for baryons is summarized below.

A meson in large-$N_c$ QCD is created with unit amplitude by the operator
\begin{equation} 
{1 \over \sqrt{N_c}} \sum_{i=1}^{N_c} \bar q_i q^i \ , 
\end{equation}
where $i$ is the color index of the quark.  The $N_c$-dependence of meson amplitudes can 
be obtained by studying
quark-gluon diagrams.  The leading diagrams are planar diagrams with a single quark loop
($L=1$) with all insertions of meson operators on the quark loop.  An $3$-meson diagram
is shown in Fig.~3 as an example.  The leading diagrams for an $n$-meson amplitude are 
$O( N_c^{1-n/2})$.  For example, a meson decay constant is $O(\sqrt{N_c})$; a meson mass is $O(1)$;
a 3-meson coupling is $O( 1/ \sqrt{N_c} )$, and so forth.  This power counting
implies that large-$N_c$ mesons are narrow states which are weakly coupled to one
another~\cite{ven, witten}.

\begin{figure}\label{threemeson}
\resizebox{.25\textwidth}{!}
{\includegraphics{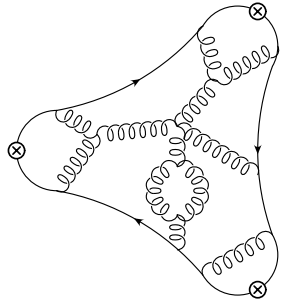}}
\caption{A planar diagram contributing to a meson three-point vertex
at leading order.}
\end{figure}

The situation for baryons is qualitatively different.  A large-$N_c$ baryon is a bound state of
$N_c$ valence quarks completely antisymmetrized in the color indices of the quarks,
\begin{equation}
\epsilon_{i_1 i_2 i_3\cdots i_{N_c}}\ q^{i_1} q^{i_2} q^{i_3} 
\cdots q^{i_{N_c}}\ .
\end{equation}
The mass of the baryon is $O(N_c)$, whereas the size of the baryon is $O(1)$.  The
$N_c$-dependence of baryon-meson scattering amplitudes and couplings can be determined by
studying the $N_c$-counting of quark-gluon diagrams.  An antibaryon-baryon--$n$-meson vertex
is $O(N_c^{1 -n/2})$, as is an amplitude for the scattering baryon+meson
$\rightarrow$ baryon + $(n-1)$ mesons.  

Naively, this power counting 
is inconsistent. 
The amplitude
for baryon + meson $\rightarrow$ baryon + meson scattering is $O(1)$, whereas an 
antibaryon-baryon--meson vertex is $O(\sqrt{N_c})$, and grows with $N_c$.  
A tree 
diagram with two different single-meson--baryon vertices produces an amplitude which is 
$O(N_c)$, not $O(1)$.  Unless the $O(N_c)$ contributions of
different tree diagrams all cancel one another exactly, the total scattering amplitude is
$O(N_c)$ and will 
violate the $1/N_c$ power counting.  Imposing the constraint that the scattering ampitude be 
$O(1)$ results in relations amongst  
single-meson--baryon-antibaryon vertices which must be satisfied for consistency of the 
$1/N_c$ power counting for baryon-meson scattering amplitudes.  
Consistency of 
$1/N_c$ power counting for baryon-meson
scattering amplitudes and vertices results in non-trivial constraints on 
large-$N_c$ baryon matrix elements at leading and subleading orders~\cite{dm,j}.    
Large-$N_c$ consistency conditions also lead to the derivation of 
contracted spin-flavor symmetry for baryons~\cite{dm,gs}.

\section{Spin-Flavor Symmetry of Large-$N_c$ Baryons}

Large-$N_c$ contracted spin-flavor symmetry can be derived by considering pion-baryon 
scattering at low energies $E \sim O(1)$.  In this kinematic regime, the large-$N_c$ baryon 
acts as a heavy static
source for scattering the pion with no recoil.  There are two tree diagrams which
contribute to the scattering amplitude at $O(N_c)$, the direct and crossed diagrams.   
Using the $N_c$-independent baryon propagator of Heavy Baryon Chiral 
Perturbation Theory~\cite{bchpt},
it is easy to show that cancellation of the $O(N_c)$ scattering amplitude from these two
diagrams is given by the large-$N_c$ consistency condition \cite{dm}
\begin{equation}\label{xx}
N_c \left[ X^{ia}, X^{jb} \right] \le O(1)\ ,  
\end{equation}
where the baryon axial vector couplings (in the baryon rest frame) are defined by 
\begin{equation}
A^{i a} \equiv g N_c X^{ia}\ .
\end{equation}
Expanding the operator $X^{ia}$ in a power series in $1/N_c$,
\begin{equation}
X^{ia} = X_0^{ia} + {1 \over N_c} X_1^{ia} + {1 \over N_c^2} X_2^{ia} +
\ldots \ ,
\end{equation}
and substituting into Eq.~(\ref{xx}) yields the constraint
\begin{equation}\label{x0x0}
\left[ X_0^{ia}, X_0^{jb} \right] = 0 
\end{equation}
for the leading $O(N_c)$ matrix elements of the baryon axial vector couplings.
As we will see, the 
matrix elements of $X_0^{ia}$ between different baryon states are all
determined relative to one another by this constraint, so the $O(N_c)$ portion of the 
baryon axial vector couplings $A^{ia}$ are all related by symmetry up to an overall
normalization constant given by the coupling $g$, which is 
the reduced matrix element of the axial vector couplings.

The operator $X_0^{ia}$ is an
irreducible tensor operator transforming according to the 
spin-$1$, $SU(3)$ adjoint representation of spin $\otimes$ flavor, so 
the commutators of $X_0^{ia}$ with the baryon spin and flavor generators 
are given by
\begin{equation}\label{jxtx}
\left[J^i, X_0^{ja} \right] = i \epsilon^{ijk} X_0^{ka}, \qquad
\left[T^a, X_0^{ib} \right] = i f^{abc} X_0^{ic} \ . 
\end{equation}      
The Lie algebra of the baryon spin $\otimes$ flavor generators, together with 
Eqs.~(\ref{x0x0}) and~(\ref{jxtx}), yields a contracted spin-flavor algebra \cite{dm, gs}
\begin{eqnarray}
&&\left[J^i, J^j \right] = i \epsilon^{ijk} J^k, 
\quad\left[T^a, T^b \right] = i f^{abc} T^c,
\quad\left[J^i, I^a \right] = 0, \nonumber\\
&&\left[J^i, X_0^{ja} \right] = i \epsilon^{ijk} X_0^{ka}, \qquad
\left[T^a, X_0^{ib} \right] = i f^{abc} X_0^{ic} \ \\ 
&&\left[ X_0^{ia}, X_0^{jb} \right] = 0  \  \nonumber
\end{eqnarray}
in the large-$N_c$ limit.

It is instructive to contrast the contracted spin-flavor algebra with the $SU(6)$ spin-flavor
algebra
\begin{eqnarray}
&&\left[J^i, J^j \right] = i \epsilon^{ijk} J^k, 
\quad\left[T^a, T^b \right] = i f^{abc} T^c,
\quad\left[J^i, T^a \right] = 0,\nonumber\\ 
&&\left[J^i, G^{ja} \right] = i \epsilon^{ijk} G^{ka}, \qquad
\left[T^a, G^{ib} \right] = i f^{abc} G^{ic}, \\
&&\left[G^{ia}, G^{jb} \right] = 
{i \over 6} \delta^{ab}\epsilon^{ijk} J^k +
{i \over 4} \delta^{ij}f^{abc} T^c + {i \over 2} \epsilon^{ijk}d^{abc} G^{kc} \ . \nonumber
\end{eqnarray}
The contracted spin-flavor algebra can be obtained from the $SU(6)$ spin-flavor algebra 
with the identification
\begin{equation}
\lim_{N_c \rightarrow\infty}{ G^{ia} \over N_c} \rightarrow X_0^{ia} \ . 
\end{equation}
Thus, the $SU(6)$ spin-flavor algebra correctly reproduces the contracted spin-flavor algebra
in the large-$N_c$ limit.  It differs from the contracted spin-flavor algebra by the inclusion
of some subleading $1/N_c$ terms in the generators $G^{ia}$.  

The contracted spin-flavor algebra in the large-$N_c$ limit leads to
baryon spin-flavor representations which are infinite-dimensional.  For finite $N_c$, it is
convenient to work with the $SU(6)$ spin-flavor algebra which leads to finite-dimensional 
baryon representations.  Since the emphasis of these lectures is on QCD baryons with $N_c=3$, 
the spin-flavor symmetry will be implemented for finite $N_c$.  Discussion of the connection
between finite-dimensional and infinite-dimensional baryon representations can be found in
Ref.~\cite{arnps}.

\begin{figure}
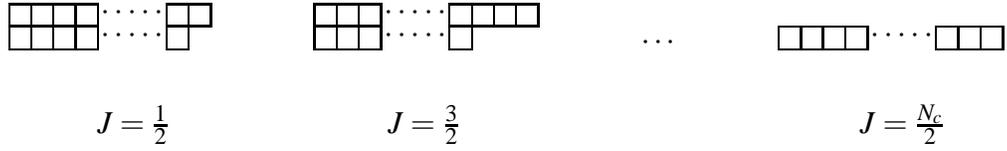
\label{baryontower}
\vbox{
\centerline{ $\nboxF$ \hskip.5in $\nboxE$ \hskip.5in $\cdots$ \hskip.5in $\nbox$ }
\vskip.25in
\centerline{ \hskip.1in ${J= \frac 1 2}$ \hskip1.1in ${J= \frac 3 2}$ \hskip2.05in ${J= \frac {N_c} 2}$}
}
\caption{Decomposition of the $SU(6)$ baryon representation $\nbox$ 
into $SU(2) \otimes SU(3)$ baryon representations.  Each Young
tableau has $N_c$ boxes.}
\end{figure}

The lowest-lying large-$N_c$ baryon representation of the spin-flavor algebra is given by
the completely symmetric tensor product of $N_c$ quarks in the fundamental rep of spin-flavor.
Under the breakdown of spin-flavor symmetry to its spin $\otimes$ flavor subgroup,
the completely symmetric spin-flavor representation decomposes into the spin and flavor
representations displayed in Fig.~4.  The baryon spin-flavor representation contains 
baryons with spins $J = \frac 1 2 , \frac 3 2, \frac 5 2, \cdots, \frac {N_c} 2$.  
The $SU(3)$ flavor representation of the baryons with a given spin
$J$ is given by the same Young tableau as its spin $SU(2)$ representation.  The dimensions of
the spin representations do not vary with $N_c$, but the dimensions of the flavor
representations do.  Consequently, the $SU(3)$ flavor multiplets are 
considerably more complicated for $N_c > 3$ than they are in QCD.  
The $(T^3, T^8)$ weight diagrams for the spin-1/2 and spin-3/2 flavor multiplets for 
large-$N_c$ baryons are given in Figs.~5 and~6.  
The numbers appearing in the weight diagrams denote the degeneracy of each weight.  While
there are many additional baryon states for $N_c > 3$, 
for $N_c=3$ these flavor representations reduce to the usual octet and decuplet multiplets,
respectively.  In the end, we will apply the $1/N_c$ expansion to QCD baryons with $N_c=3$, and
there will be no unphysical baryon states in the expansion.    

\setlength{\unitlength}{3mm}
\begin{figure}\label{spin12}
\vbox{
\centerline{\hbox{
\begin{picture}(20.79,18)(-10.395,-8)
\multiput(-1.155,10)(2.31,0){2}{\onedot}
\multiput(-2.31,8)(4.62,0){2}{\onedot}
\multiput(-3.465,6)(6.93,0){2}{\onedot}
\multiput(-4.62,4)(9.24,0){2}{\onedot}
\multiput(-5.775,2)(11.55,0){2}{\onedot}
\multiput(-6.93,0)(13.86,0){2}{\onedot}
\multiput(-8.085,-2)(16.17,0){2}{\onedot}
\multiput(-9.24,-4)(18.48,0){2}{\onedot}
\multiput(-10.395,-6)(20.79,0){2}{\onedot}
\multiput(-9.24,-8)(2.31,0){9}{\onedot}
\multiput(0,8)(2.31,0){1}{\twodot}
\multiput(-1.155,6)(2.31,0){2}{\twodot}
\multiput(-2.31,4)(2.31,0){3}{\twodot}
\multiput(-3.465,2)(2.31,0){4}{\twodot}
\multiput(-4.62,0)(2.31,0){5}{\twodot}
\multiput(-5.775,-2)(2.31,0){6}{\twodot}
\multiput(-6.93,-4)(2.31,0){7}{\twodot}
\multiput(-8.085,-6)(2.31,0){8}{\twodot}
\end{picture}
}}}
\caption{$SU(3)$ weight diagram of spin-$\frac 1 2$ baryons for large $N_c$.}
\end{figure}
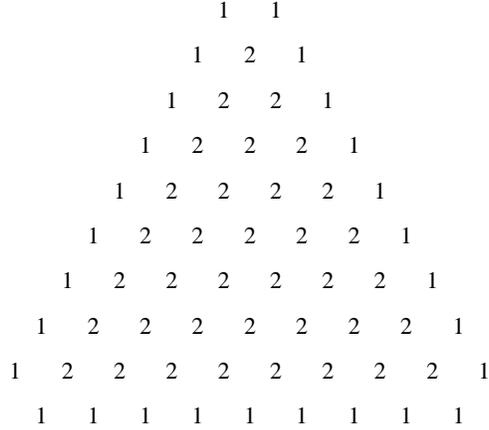

\setlength{\unitlength}{3mm}
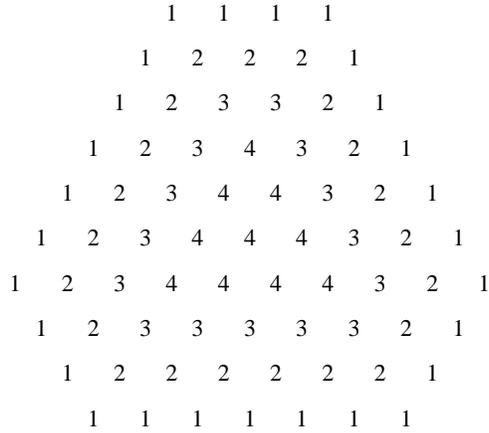
\begin{figure}\label{spin32}
\vbox{
\centerline{\hbox{
\begin{picture}(20.79,18)(-8.085,-8)
\multiput(-1.155,10)(2.31,0){4}{\onedot}
\multiput(-2.31,8)(9.24,0){2}{\onedot}
\multiput(-3.465,6)(11.55,0){2}{\onedot}
\multiput(-4.62,4)(13.86,0){2}{\onedot}
\multiput(-5.775,2)(16.17,0){2}{\onedot}
\multiput(-6.93,0)(18.48,0){2}{\onedot}
\multiput(-8.085,-2)(20.79,0){2}{\onedot}
\multiput(-6.93,-4)(18.48,0){2}{\onedot}
\multiput(-5.775,-6)(16.17,0){2}{\onedot}
\multiput(-4.62,-8)(2.31,0){7}{\onedot}
\multiput(0,8)(2.31,0){3}{\twodot}
\multiput(-1.155,6)(6.93,0){2}{\twodot}
\multiput(-2.31,4)(9.24,0){2}{\twodot}
\multiput(-3.465,2)(11.55,0){2}{\twodot}
\multiput(-4.62,0)(13.86,0){2}{\twodot}
\multiput(-5.775,-2)(16.17,0){2}{\twodot}
\multiput(-4.62,-4)(13.86,0){2}{\twodot}
\multiput(-3.465,-6)(2.31,0){6}{\twodot}
\multiput(1.155,6)(2.31,0){2}{\threedot}
\multiput(0,4)(4.62,0){2}{\threedot}
\multiput(-1.155,2)(6.93,0){2}{\threedot}
\multiput(-2.31,0)(9.24,0){2}{\threedot}
\multiput(-3.465,-2)(11.55,0){2}{\threedot}
\multiput(-2.31,-4)(2.31,0){5}{\threedot}
\multiput(2.31,4)(2.31,0){1}{\fourdot}
\multiput(1.155,2)(2.31,0){2}{\fourdot}
\multiput(0,0)(2.31,0){3}{\fourdot}
\multiput(-1.155,-2)(2.31,0){4}{\fourdot}
\end{picture}
}}}
\caption{$SU(3)$ weight diagram of the spin-$\frac 3 2$ baryons for large $N_c$.}
\end{figure}

\section{$1/N_c$ Expansion for Baryons}

The $1/N_c$ expansion of any baryon operator can be obtained by solving large-$N_c$
consistency conditions.  Each baryon operator has a $1/N_c$ expansion in terms of
all independent operator products which can be constructed from 
the baryon spin-flavor generators.  

The general form of the baryon
$1/N_c$ expansion is given by
\begin{equation}  
{\cal O}^{m}_{\rm QCD} 
= N_c^m \ \sum_{n=0}^{N_c} c_n \  { 1 \over {N_c^{n} }} {\cal O}^n \ ,
\end{equation}
where ${\cal O}^{m}_{\rm QCD}$ is an 
$m$-body quark operator in QCD which acts on baryon states.  
The baryon matrix elements of an $m$-body QCD quark operator can be at most $O(N_c^m)$, which
is reflected in the factor of $N_c^m$ in front of the $1/N_c$ operator expansion.  The $1/N_c$
expansion sums over all independent operators which transform according to the same
representation of spin $\otimes$ flavor symmetry as the QCD operator.  The independent
operators ${\cal O}^n$ which form a basis for the $1/N_c$ expansion are 
$n^{\rm th}$ degree polynomials of the baryon spin-flavor generators $J^i$, $T^a$ and $G^{ia}$.  
The baryon matrix elements of
these operator products can be computed in terms of the matrix elements of the baryon spin-flavor
generators.  The order in $1/N_c$ at which each independent operator product
appears in the $1/N_c$ expansion also is known.  $1/N_c$ power counting implies that an $n$-body
operator product is multiplied by an explicit factor of $1/N_c^n$, or that each spin-flavor
generator in an operator product is accompanied by a factor of $1/N_c$.  
Since the matrix elements of an $n$-body operator product are $\le
O(N_c^n)$, the matrix elements of each term in the $1/N_c$ operator expansion are 
manifestly $\le O(N_c^m)$, 
as required.
Every operator in the operator basis is accompanied by an unknown coefficient $c_n$,
which is a reduced matrix element of the spin-flavor $1/N_c$ expansion and is not predicted by
spin-flavor symmetry.  The coefficients are $O(1)$ at leading order in
the $1/N_c$ expansion.  Finally, the summation over spin-flavor operators $O^n$ only extends to
$N_c$-body quark operators, since all $n$-body quark operators with $n > N_c$ will be redundant
operators on baryons composed of $N_c$ valence quarks.    

Each operator
${\cal O}^n$ in the operator basis of the $1/N_c$ expansion can be written as an 
$n$-body quark operator using
bosonic quarks which carry only spin and flavor quantum numbers.  Let $q^{\alpha}$
represent an annihilation operator for a bosonic quark with spin-flavor   
$\alpha=1,\cdots, 6$, and let $q^\dagger_\alpha$ represent the corresponding creation operator.
In terms of these creation/annihilation operators, all $n$-body quark operators acting on baryon
states can be catalogued. 
There is only a single $0$-body quark operator, the baryon identity operator
$\openone$, which does not act on any of the valence quarks in the baryon.  
The $1$-body quark operators are 
given by $q^\dagger q$ and the baryon
spin-flavor generators
\begin{eqnarray}  
J^i &=& q^\dagger \left({ \sigma^i \over 2} \otimes 
\openone \right) q \ , \nonumber\\
T^a &=& q^\dagger \left(\openone \otimes 
{\lambda^a \over 2} \right) q \ , \\
G^{ia} &=& q^\dagger \left({\sigma^i \over 2} 
\otimes {\lambda^a \over 2}\right) q \ .\nonumber
\end{eqnarray}
The notation here is compact.  Each $1$-body quark operator is understood to act on each quark
line in the baryon.  Thus,
\begin{equation}
G^{ia} = \sum_\ell q_\ell^\dagger \left({\sigma^i \over 2} 
\otimes {\lambda^a \over 2}\right) q_\ell \ .
\end{equation}
In addition, note that $q^\dagger q = N_c \openone$ is not an independent operator and 
can be eliminated from the list of $1$-body quark operators.
The $2$-body quark operators are given by products of the $1$-body quark operators, the baryon
spin-flavor generators.  Each $2$-body operator product can be written as the symmetric product
(or anticommutator) of two $1$-body operators since the 
commutator of any two $1$-body operators can be replaced by a linear
combination of $1$-body operators by the spin-flavor algebra.  This observation also applies to
all $n$-body operators with $n \ge 2$.

\begin{table}
\begin{tabular}{cc}
\hline
\tablehead{1}{c}{b}{Identity}
&\tablehead{1}{l}{b}{( J, SU(3) )} \\
\hline
\smallskip
$2\ \left\{J^i,J^i\right\} + 3\ \left\{T^a,T^a\right\} + 12\ 
\left\{G^{ia},G^{ia}\right\} = 5 N_c \left(N_c+6\right)$&$(0,0)$\\
\hline
\smallskip
$d^{abc}\ \left\{G^{ia}, G^{ib}\right\} + {2\over 3}\ \left\{J^i,G^{ic}
\right\} + {1\over4}\ d^{abc}\ \left\{T^a, T^b\right\} = {2\over 3}
\left(N_c+3\right)\ T^c $ & $(0,8)$ \\
\smallskip
$\left\{T^a,G^{ia}\right\} = {2\over3}\left(N_c+3\right)\ J^i$ & $(1,0)$\\
\smallskip
${1\over 3}\ \left\{J^k,T^c\right\} +  d^{abc}\ \left\{T^a,G^{kb}\right\}
-\epsilon^{ijk} f^{abc} \left\{G^{ia}, G^{jb}\right\} =  {4\over3}
\left(N_c+3\right)\ G^{kc}$ & $(1,8)$\\
\hline
\smallskip
$-12\ \left\{G^{ia},G^{ia}\right\} + 27\ \left\{T^a,
T^a\right\} - 32\ \left\{J^i,J^i\right\}=0 $& $(0,0)$\\
\smallskip
$d^{abc}\ \left\{G^{ia}, G^{ib}\right\} + {9\over 4} \ d^{abc}\ \left\{
T^a, T^b\right\} - {10\over3}\ \left\{J^i,G^{ic}\right\} = 0 $& $(0,8)$\\
\smallskip
$4\ \left\{G^{ia},G^{ib}\right\} = \left\{T^a,T^b\right\}\qquad ({27})$
& $(0,{27})$\\
\smallskip
$\epsilon^{ijk}\ \left\{ J^i,G^{jc}\right\} = f^{abc} \ \left\{T^a,G^{kb}
\right\}$& $(1,8)$\\
\smallskip
$3\ d^{abc}\ \left\{T^a,G^{kb}\right\} = \left\{J^k,T^c\right\} -  
\epsilon^{ijk} f^{abc}\ \left\{G^{ia}, G^{jb}\right\}$& $(1,8)$\\
\smallskip
$\epsilon^{ijk}\ \left\{G^{ia},G^{jb}\right\} = f^{acg} d^{bch}\ \left\{
T^g,G^{kh}\right\}\qquad ({10}+{\overline {10}})$ & 
$(1,{10}+{\overline {10}})$\\
\smallskip
$3\ \left\{G^{ia}, G^{ja}\right\} = \left\{J^i, J^j
\right\}\qquad (J=2)$ & $(2,0)$\\
\smallskip
$3\ d^{abc}\ \left\{G^{ia}, G^{jb}\right\} = 
\left\{J^i,G^{jc}\right\}\qquad (J=2)$ & $(2,8)$\\
\hline
\end{tabular}
\caption{$SU(6)$ Operator Identities}
\label{tab:a}
\end{table}

Not all operator products of the spin-flavor generators are linearly dependent, so it is
necessary to eliminate redundant operators using operator identities.  The complete set of
$2$-body operator identities for the completely symmetric baryon representation of $SU(6)$ 
are given in Table~1 along with their respective spin $\otimes$ flavor representations
\cite{djm2}.  It is possible to
eliminate all redundant $n$-body operators by using the $2$-body operator identities, so
Table~1 gives the complete set of operator identities. 
  
The group theory behind
Table~1 is interesting.  Purely $n$-body quark operators are normal-ordered operators of
the form
\begin{equation}
q^\dagger_{\alpha_1} \ldots q^\dagger_{\alpha_n}\  
{T}^{\alpha_1 \cdots \alpha_n}_{\beta_1 
\cdots \beta_n}\  
q^{\beta_1} \ldots q^{\beta_n}\ .
\end{equation} 
For the completely symmetric $SU(6)$ representation of baryon states, the only nonvanishing
normal-ordered quarks operators have tensors $T$ which are totally symmetric in the upper and
the lower spin-flavor
indices.  We would like to reexpress these independent normal-ordered operators
as operator products of the spin-flavor generators, whose baryon matrix elements are known.

The group theory behind the operator identities is particularly elegant.  The $0$-body quark
operator $\openone$ is in the singlet representation of $SU(6)$, whereas  
the $1$-body operators transform as the tensor product of a fundamental and antifundamental of
$SU(6)$, which decomposes into a singlet and an adjoint of $SU(6)$.  The $2$-body operators
are obtained from the tensor product of the symmetric $2$-quark representation and its conjugate. 
Specifically,
\begin{eqnarray}
&&{\rm 0-body:}\ \ 1 \nonumber\\
&&{\rm 1-body:}\ \ \left( \overline{\onebox} \otimes \onebox \right) = 1 + {\rm adj} 
= 1 + T^\alpha_\beta \\
&&{\rm 2-body:}\ \left( \overline{\twobox} \otimes \twobox \right) 
 = 1 + T^\alpha_\beta + T^{(\alpha_1 \alpha_2)}_{(\beta_1 \beta_2)}. \nonumber
\end{eqnarray}
There are identities which relate the singlet and adjoint $2$-body operators to
$1$-body adjoint and $0$-body singlet operators, so the $2$-body operators which transform as
the singlet and the adjoint are not independent.  The relevant identities 
are most easily understood keeping $SU(6)$ symmetry manifest.  
The generators $J^i$, $T^a$ and $G^{ia}$ form a complete set of $SU(6)$ generators $\Lambda^A$, 
$A=1, \cdots, 35$.  The operator identities relating the singlet $2$-body operators to the
$0$-body operator and the adjoint $2$-body operators to the $1$-body operators
are given by the Casimir identities
\begin{eqnarray}
&&\Lambda^A \Lambda^A = C(R) \ \openone \nonumber\\
&&d^{ABC}\Lambda^B \Lambda^C = D(R)\ \Lambda^A , 
\end{eqnarray}
where $C(R)$ and $D(R)$ are the quadratic and cubic Casimirs for the $SU(6)$ baryon
representation $R$. 
These Casimir identities for the completely symmetric
baryon spin-flavor representation produce the operator identities in the first two blocks of
Table~1.  The remaining $2$-body operator identities arise because the completely symmetric
product of two $SU(6)$ adjoints,
\begin{equation} 
({\rm adj} \otimes {\rm adj})_S =1 + T^\alpha_\beta + T^{[\alpha_1 \alpha_2]}_{[\beta_1 \beta_2]}
+ T^{(\alpha_1 \alpha_2)}_{(\beta_1 \beta_2)} \ ,
\end{equation}
contains an additional tensor structure.
The $2$-body operator products corresponding to
the tensor $T^{[\alpha_1 \alpha_2]}_{[\beta_1 \beta_2]}$ will vanish identically
when acting on the completely symmetric baryon spin-flavor representation.  
These operator product combinations yield the vanishing operator identities
given in the third part of Table~1.

The above operator identities are summarized by the following operator reduction rule:
All operators in which 
two flavor indices are contracted using
$\delta^{ab}$, $d^{abc}$, or $f^{abc}$ or two spin indices on $G$'s are
contracted using $\delta^{ij}$ or $\epsilon^{ijk}$ can be eliminated.

\section{QCD Baryons}

I will now derive $1/N_c$ expansions
for the masses, axial vector currents and magnetic moments of baryons in QCD.  
$SU(3)$ flavor breaking
cannot be neglected relative to $1/N_c$, and 
is included in the analysis.  For large-$N_c$ baryons, the $1/N_c$ expansion extends up to 
$N_c$-body operators, so the $1/N_c$ expansion for QCD baryons goes up to 
third order in the generators.    
The $1/N_c$ expansion including flavor symmetry breaking
also goes up to $N_c$-body operators, so perturbative $SU(3)$ breaking 
extends to finite order in flavor symmetry breaking.  For a baryon operator with a $1/N_c$ expansion
beginning with a $n$-body operator, the flavor symmetry breaking expansion extends
to order $(N_c -n)$.

\subsection{Masses}

The baryon mass operator is a $J=0$ operator.  The leading operator in the $1/N_c$ expansion
is the flavor singlet operator $N_c \openone$ which gives the same $O(N_c)$ mass to all
baryons in a spin-flavor representation.  
Since the $1/N_c$ expansion begins with a $0$-body operator, the baryon mass operator
can be expanded to third order in flavor symmetry breaking.  Thus, the baryon mass operator
decomposes into the $SU(3)$ flavor representations
\begin{equation}\label{massop}
M = M^{\bf 1} + M^{\bf 8} + M^{\bf 27} + M^{\bf 64},
\end{equation}
where the singlet, octet, ${\bf 27}$ and ${\bf 64}$ are zeroth, first, second and third order
in $SU(3)$ flavor symmetry breaking, respectively.  Each of these spin-singlet flavor
representations has a $1/N_c$ operator expansion.  The $1/N_c$ expansions are given by
\begin{eqnarray}\label{mop}
&&M^{\bf 1} = N_c \openone + {1 \over N_c}J^2, \nonumber\\
&&M^{\bf 8} = T^8 + {1 \over N_c} \left\{J^i, G^{i8} \right\}
+ {1 \over N_c^2} \left\{ J^2, T^8 \right\}, \nonumber\\
&&M^{\bf 27} = {1 \over N_c}\left\{ T^8, T^8 \right\} 
+{1 \over N_c^2}\left\{ T^8, \left\{ J^i, G^{i8} \right\}\right\}, \\
&&M^{\bf 64} = {1 \over N_c^2}\left\{ T^8, \left\{ T^8, T^8
\right\}\right\}\ , \nonumber
\end{eqnarray}
where it is to be understood that there is an unknown coefficient multiplying each operator 
in the above $1/N_c$ expansions.  Note that the operators in these expansions can be derived 
using the operator identities in Table~1.  For example, consider the octet mass expansion.
$SU(3)$ flavor breaking transforms as the eighth component of an octet.  There is only one
$1$-body operator which is $J=0$ and the eighth component of an $SU(3)$ octet, namely $T^8$.
From Table~1, one finds that there are three $2$-body operators which transform in this manner:
$d^{ab8} \left\{G^{ia}, G^{ib} \right\}$, $d^{ab8} \left\{ T^a, T^b \right\}$ and 
$\left\{J^i, G^{i8} \right\}$.  However, Table~1 shows that one linear combination of these
operators is proportional to the $1$-body operator $T^8$, and that another linear combination
vanishes for the completely symmetric $SU(6)$ baryon representation.  Thus, there is only
one independent $2$-body operator.  This $2$-body operator is taken to be 
$\left\{J^i, G^{i8} \right\}$ by the operator reduction rule.  Application of the $2$-body
identities implies that there is a single independent $3$-body operator which transforms as a
spin singlet and as the eighth component of a flavor octet.  Without loss of generality, this
operator can be taken to be $\left\{ J^2, T^8 \right\}$.  Similar analyses produce the other
expansions in Eq.~(\ref{mop}).

\begin{table}
\begin{tabular}{cccc}
\hline
\tablehead{1}{c}{b}{Mass Splitting}
&\tablehead{1}{c}{b}{$1/N_c$}
&\tablehead{1}{c}{b}{Flavor}
&\tablehead{1}{c}{b}{Expt.} \\
\hline
\smallskip
${5 \over 8}(2N +3\Sigma
+\Lambda +2\Xi) -{1 \over {10}}(4\Delta +3\Sigma^* +2\Xi^* +\Omega)$ & $N_c$
& $1$ & * \\
\smallskip
${1 \over 8}(2N+ 3\Sigma +\Lambda
+2\Xi) -{1 \over {10}}(4\Delta +3\Sigma^* +2\Xi^* +\Omega)$ & $1/N_c$ & $1$
& $18.21 \pm 0.03\%$ \\
\smallskip
${5 \over 2}(6N -3\Sigma +\Lambda
-4\Xi) -(2\Delta -\Xi^* -\Omega)$ & $1$ & $\epsilon$ &
$20.21 \pm 0.02\%$ \\
\smallskip
${1 \over 3}(N -3\Sigma +\Lambda
+\Xi)$ & $1/N_c$ & $\epsilon$ & $5.94 \pm 0.01\%$ \\
\smallskip
${1 \over 2}(-2N -9\Sigma
+3\Lambda + 8\Xi) +(2\Delta -\Xi^* -\Omega)$ & $1/N_c^2$ &
$\epsilon$ & $1.11 \pm 0.02\%$ \\
\smallskip
${5 \over 4}(2N -\Sigma
-3\Lambda +2\Xi) -{1 \over 7}(4\Delta -5\Sigma^* -2\Xi^* +3\Omega)$ &
$1/N_c$ & $\epsilon^2$ & $0.37 \pm 0.01\%$ \\
\smallskip
${1 \over 2} (2N -\Sigma
-3\Lambda + 2\Xi) -{1 \over 7}(4\Delta -5\Sigma^* -2\Xi^* +3\Omega)$ &
$1/N_c^2$ & $\epsilon^2$ & $0.17 \pm 0.02\%$ \\
\smallskip
${1 \over 4}(\Delta - 3 \Sigma^* + 3
\Xi^* - \Omega)$ & $1/N_c^2$ & $\epsilon^3$ & $0.09 \pm 0.03\%$ \\
\hline
\end{tabular}
\caption{Baryon Mass Hierarchy}
\label{tab:two}
\end{table}

\begin{figure}\label{masses}
\resizebox{1.0\textwidth}{!}
{\includegraphics{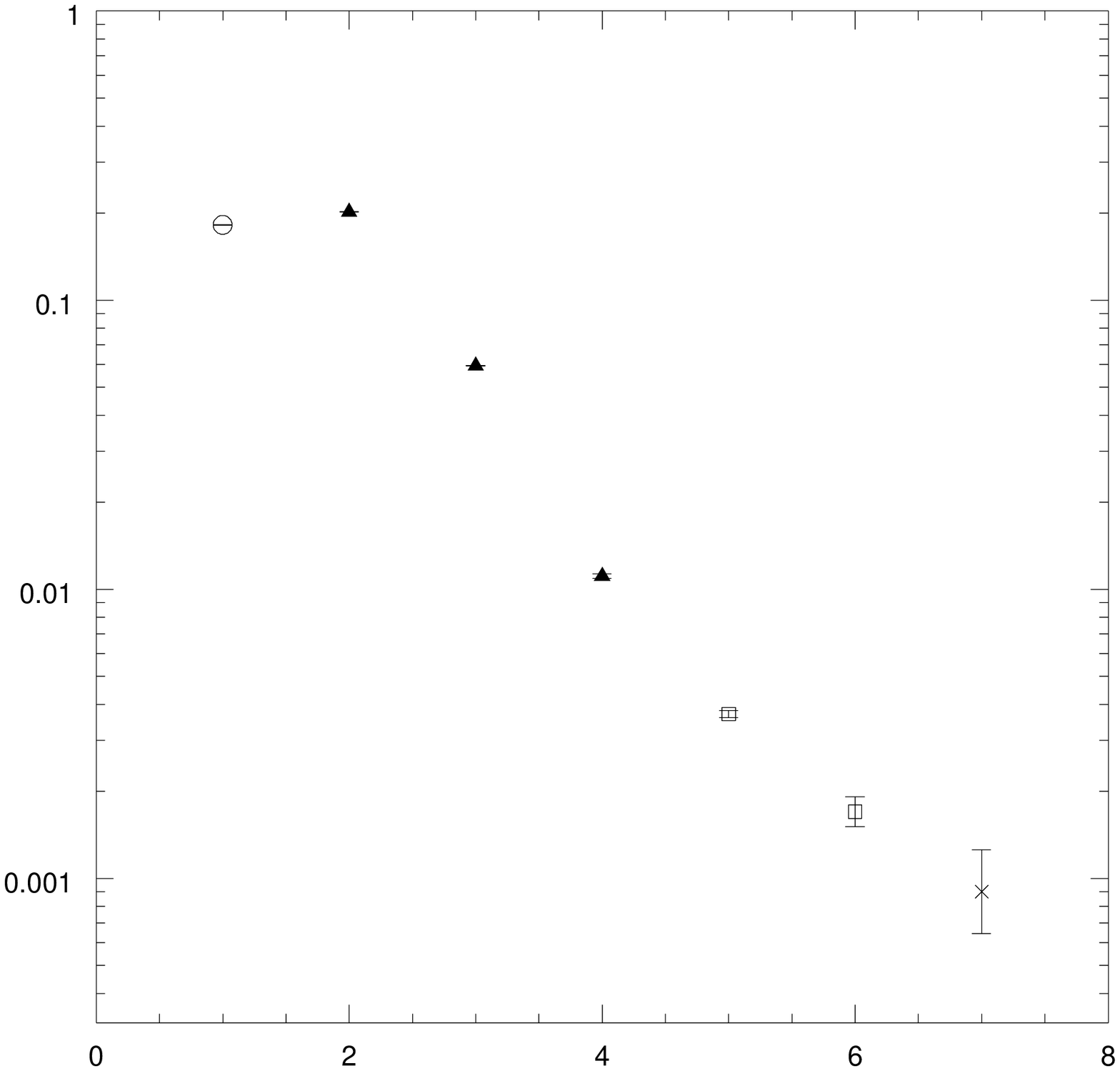}}
\caption{Baryon mass hierarchy.  The mass combinations are of relative order
${1 \over N_c^2}$, ${\epsilon \over N_c}$, ${\epsilon \over N_c^2}$,
${\epsilon \over N_c^3}$, ${\epsilon^2 \over N_c^2}$, ${\epsilon^2 \over N_c^3}$,
${\epsilon^3 \over N_c^3}$ compared to the overall $O(N_c)$ singlet mass of the baryon
$\bf 56$.
}
\end{figure}

The $1/N_c$ expansion of the baryon mass operator given by Eqs.~(\ref{massop}) and~(\ref{mop})
contains eight independent operators, which is equal to the number of baryon masses in the
$\bf 56$ of $SU(6)$:
the $N$, $\Lambda$, $\Sigma$, $\Xi$, $\Delta$, $\Sigma^*$, $\Xi^*$, $\Omega$.
Each mass operator contributes to a unique linear combination of these eight masses.  These
linear combinations are given in Table~2\cite{jl}.  Each mass combination occurs at specific orders in the
$1/N_c$ and $SU(3)$ flavor breaking expansions;  
these suppression factors also appear in Table~2.
The parameter $\epsilon \sim
m_s/\Lambda_{\rm QCD}$ is the suppression factor for $SU(3)$ flavor symmetry breaking.  The
final column gives the experimental value for the accuracy of each mass combination, which is
defined by the dimensionless quantity 
\begin{equation}
{{\sum B_i} \over {\sum |B_i|/2}}
\end{equation}
computed for each mass combination.  

The $1/N_c$ and flavor symmetry breaking hierarchy
predicted for the baryon masses can be tested by comparing the experimental accuracies to
the $1/N_c$ and $\epsilon$ suppression factors.  
The numerical accuracies of the mass combinations are plotted in Fig.~7, except for 
the mass combination corresponding to the $N_c \openone$ operator.
The hierarchy predicted by the $1/N_c$ suppression factors is clearly evident.  For example,
there are three mass combinations that are first order in $SU(3)$ breaking, but of 
order $1/N_c$, $1/N_c^2$ and $1/N_c^3$ relative to the leading $O(N_c)$ singlet mass
of the baryons.  
This pattern can be seen in Fig.~7.  In addition,
the two flavor $\bf 27$ mass combinations which are second order in $SU(3)$ breaking
are suppressed by factors of $1/N_c^2$ and $1/N_c^3$ relative to the leading $O(N_c)$ baryon
mass.  The Gell-Mann--Okubo flavor-$\bf 27$ 
mass splitting of the spin-1/2 baryon octet,
\begin{equation}
\frac 1 4 \left( 2N - \Sigma - 3 \Lambda + 2\Xi \right) ,
\end{equation}
and the flavor-$\bf 27$ Equal Spacing Rule mass splitting of the spin-3/2 baryon decuplet,
\begin{equation}
\frac 1 7 \left( 4 \Delta - 5 \Sigma^* - 2\Xi^* + 3 \Omega \right) ,
\end{equation}
are linear combinations of the two flavor-$\bf 27$ mass splittings specified by the $1/N_c$
expansion, so each is predicted to be a factor of $1/N_c^2$ more accurate than expected from
flavor symmetry breaking factors alone.  The most suppressed mass splitting is the flavor-$\bf
64$ Equal Spacing Rule mass splitting,
\begin{equation}
\frac 1 4 \left( \Delta - 3 \Sigma^* + 3 \Xi^* - \Omega \right),
\end{equation}   
which is third order in $SU(3)$ flavor breaking and of relative order $1/N_c^3$.  This mass
combination is clearly suppressed by a greater factor than predicted from $SU(3)$ breaking
alone.  The experimental accuracy of this mass combination
is consistent with the $1/N_c$ hierarchy, but a better measurement of
the splitting in needed to test the $1/N_c^3$ prediction of the $1/N_c$ expansion 
definitively.

In summary, the $1/N_c$ hierarchy is observed in the $I=0$ baryon mass splittings, and 
the presence of $1/N_c$
suppression factors explains the accuracy of baryon mass combinations quantitatively.

There also is clear evidence for the $1/N_c$ hierarchy in the $I=1$ baryon mass splittings.
For example, the Coleman-Glashow mass splitting
\begin{equation}
\left[ \left( p-n \right) - \left( \Sigma^+ - \Sigma^- \right) + \left( \Xi^0 - \Xi^- \right)
\right]
\end{equation}
has been measured to be non-zero for the first time quite recently.  
The measured mass splitting is more accurate than the 
prediction based on flavor suppression factors alone, and is consistent with an additional
$1/N_c^2$ suppression predicted by the $1/N_c$ expansion\cite{jl}.  It is particularly noteworthy that
this prediction of the $1/N_c$ expansion was made before the Coleman-Glashow mass splitting
was measured to the precision required to test the $1/N_c$ hierarchy.  A more
in-depth discussion of the $I=1$ baryon mass splittings can be found in 
Refs.~\cite{cgmass} and~\cite{lat2000}.

\subsection{Axial Vector Couplings}

The baryon axial vector current operator
$A^{ia}$ is $J=1$ and an $SU(3)$ flavor adjoint.  In the
$SU(3)$ symmetry limit, the $1/N_c$ expansion of the baryon axial vector current
is given by\cite{djm2}  
\begin{eqnarray}\label{aia}
A^{ia} = &&a_1 G^{ia} + b_2 {1 \over N_c} J^i T^a 
+ b_3 {1 \over N_c^2} \left\{ J^i, \left\{ J^j, G^{ja} \right\} \right\} \nonumber\\
&&+ d_3 {1 \over N_c^2} \left( \left\{ J^2, G^{ia} \right\} - \frac 1 2
\left\{J^i,\left\{ J^j, G^{ja} \right\} \right\} \right) \ , 
\end{eqnarray}
where the $1/N_c$ expansion for QCD baryons extends up to $3$-body operators.
The $1/N_c$ expansion involves four independent operators, so the baryon axial
vector couplings are
determined in terms of the four unknown coefficients $a_1$, $b_2$, $b_3$ and $d_3$.
The usual $SU(3)$ flavor analysis of the axial vector couplings of the spin-$1/2$
baryon octet and spin-$3/2$ decuplet is given in terms of the four $SU(3)$ couplings
$D$, $F$, $C$ and $H$,  
\begin{equation}
2 D\ \Tr\ \bar B S^\mu \left\{ {\cal A}_\mu, B \right\}
+2 F\ \Tr\ \bar B S^\mu \left[ {\cal A}_\mu, B \right]
+ C\ \left( \bar T^\mu {\cal A}_\mu B + \bar B {\cal A}_\mu T^\mu \right)
+ 2 H\ \bar T^\mu S^\nu {\cal A}_\nu T_\mu \ ,
\end{equation}
where $B$ represents the baryon octet, $T^\mu$ denotes the baryon decuplet, $S^\mu$
is a spin operator, and ${\cal A}^\mu$ is the axial vector current of the pion octet.
The coefficients of the $1/N_c$ parametrization and the $SU(3)$ couplings are related
by
\begin{eqnarray}\label{dfch}
&&D = \frac 1 2 a_1 + \frac 1 6 b_3, \nonumber\\
&&F = \frac 1 3 a_1 + \frac 1 6 b_2 + \frac 1 9 b_3, \nonumber\\
&&C = -a_1 - \frac 1 2 d_3, \\
&&H = -\frac 3 2 a_1 - \frac 3 2 b_2 -\frac 5 2 b_3 . \nonumber
\end{eqnarray}

The $1/N_c$ expansion for the baryon axial vector current can be truncated after the
first two operators in Eq.~(\ref{aia}) since the two $3$-body operators are both suppressed relative to
the $1$-body operator by a factor of $1/N_c^2$.  The $2$-body operator $J^i T^a$ can 
not be neglected relative to the $1$-body operator $G^{ia}$ for all $a=1,\cdots,8$,
so the leading order result for $A^{ia}$ is given by
\begin{equation}
A^{ia} = a_1 G^{ia} + b_2 {1 \over N_c} J^i T^a ,
\end{equation}
in terms of two parameters $a_1$ and $b_2$.  The operator $G^{ia}$ alone produces axial
vector couplings with $SU(6)$ symmetry.  If only the $G^{ia}$ operator is retained, 
the four $SU(3)$ couplings satisfy
\begin{equation}
F/D = 2/3, \qquad C= -2D, \qquad H = -3F\ .
\end{equation}
The operator $J^i T^a$ breaks the $SU(6)$
symmetry.  The breaking is such that the $SU(3)$ couplings are related by  
\begin{equation}
C = -2 D , \qquad\qquad H=3D-9F \ ,
\end{equation}
which reduces to $SU(6)$ symmetry when $F/D = 2/3$.  

It is worthwhile to consider the
isospin decomposition of $A^{ia}$ into the isovector, isodoublet and
isosinglet axial vector currents.  The $1/N_c$ expansion for the isovector axial
vector current is given by
\begin{equation}
A^{ia} = a_1 G^{ia} + b_2 {1 \over N_c} J^i I^a, \qquad a=1,2,3 \ .
\end{equation}
The $2$-body operator $J^i I^a$ is suppressed relative to $G^{ia}$ for
$a=1,2,3$, since the matrix elements of $G^{ia}$ are $O(N_c)$ whereas the matrix elements
of $J^i$ and $I^a$ are both $O(1)$.  Thus, the $1/N_c$ expansion for the isovector current can
be truncated to the $1$-body operator $G^{i3}$ up to a correction of order $1/N_c^2$ relative
to the leading $O(N_c)$ term.
A fit to the pion couplings yields $F/D = 2/3$ up to a correction
of relative order $1/N_c^2$.  Similar reasoning for the isodoublet axial vector
current gives $F/D = 2/3$ up to a correction of relative order $1/N_c$.  The $2$-body
operator cannot be neglected relative to the $1$-body operator $G^{ia}$ for the
isosinglet axial vector current with $a=8$.

The $1/N_c$ operator expansion for the flavor-octet baryon axial vector current
can be generalized to include $SU(3)$ breaking.  The $SU(3)$-symmetric expansion
begins with a $1$-body operator, so the baryon axial vector currents need to be
expanded to second order in $SU(3)$ symmetry breaking.  However, many of the baryon
axial vector current observables are not measured, so it is not necessary to
construct the $1/N_c$ expansion to all orders in $SU(3)$ breaking.  Instead,
the $1/N_c$ expansion will be considered to linear order in $SU(3)$ breaking.  

The expansion at linear order in $SU(3)$ breaking involves
additional spin-1 operators in different flavor representations,
\begin{equation}
\delta A^{ia} = A^{ia}_{\bf 1} + A^{ia}_{\bf 8_S} + A^{ia}_{\bf 8_A} +A^{ia}_{\bf 27} 
+A^{ia}_{\bf 10 + \bar {10}}
\ .
\end{equation}
A valid truncation of the $1/N_c$ expansion to first order in $SU(3)$ breaking
was constructed in Ref.~\cite{djm2}.  The $1/N_c$ expansion is given by    
\begin{eqnarray}
A^{ia} &&= \left( a_1 \delta^{ab} + c_1 d^{ab8} \right) G^{ib} 
+ \left( b_2 \delta^{ab} + c_2 d^{ab8} \right) {1 \over N_c} J^i T^b \nonumber\\
&&+c_3 {1 \over N_c} \left\{ G^{ia}, N_s \right\} + c_4 {1 \over
N_c} \left\{ J_s^i, T^a \right\} \\
&&+ \frac 1 3 c_5 {1 \over N_c} \left[ J^2,
\left[ N_s, G^{ia} \right] \right] + \frac 1 3 \left( c_1 + c_2 \right)
\delta^{a8} J^i, \nonumber
\end{eqnarray}
where the coefficients $a_1$ and $b_2$ are zeroth order in $SU(3)$ breaking, and
the coefficients $c_1, \cdots, c_5$ are first order in the $SU(3)$ breaking.  Thus,
it is to be understood that the $c_i$ are proportional to $\epsilon$.  Dropping the
$c_5$ operator (since it does not contribute to any of the measured axial couplings)
and adding the $d_3$ operator (to allow the $SU(3)$ parameters $D$, $F$ and $C$ to 
have arbitrary values) results in the 7-parameter formula 
\begin{eqnarray}\label{fitform}
A^{ia} = && a_1 G^{ia} + b_2 {1 \over N_c} J^i T^a 
+d_3 {1 \over N_c^2} \left( \left\{ J^2, G^{ia} \right\} 
- \frac 1 2 \left\{ J^i, \left\{ J^j , G^{ja} \right\} \right\} \right)\\
&&+ \Delta^a \left( c_1 G^{ia} + c_2 {1 \over N_c} J^i T^a \right)
+ c_3 {1 \over N_c} \left\{ G^{ia}, N_s \right\} + c_4 {1 \over N_c} \left\{ T^a, J_s^i \right\}
+ \frac 1 {\sqrt{3}} \delta^{a8} W^i  , \nonumber
\nonumber
\end{eqnarray}
where $\Delta_a=1$ for $a=4,5,6,7$ and is zero otherwise, and
\begin{equation}\label{w}
W^i = \left( c_4 - 2 c_1 \right) J_s^i + {1 \over N_c}
\left( c_3 - 2 c_2 \right) N_s J^i - 3{1 \over N_c}
\left( c_3 + c_4 \right) N_s J_s^i .
\end{equation}

\begin{table}
\begin{tabular}{cc}
\hline
& Fit A \\ 
\hline
$a_1$   & 1.764 $\pm $ 0.042 \\
$b_2$   &-1.218 $\pm $ 0.216 \\
$d_3$   &0.549 $\pm $ 0.081 \\
$c_1$ &-0.044 $\pm $ 0.048 \\
$c_2$ & 0.792 $\pm $ 0.228 \\
$c_3$ &-0.432 $\pm $ 0.036 \\
$c_4$ & 0.096 $\pm $ 0.072 \\
$F$   & 0.39  $\pm $ 0.02  \\
$D$   & 0.88  $\pm $ 0.02  \\
$3F-D$& 0.27  $\pm $ 0.09  
\end{tabular}
\caption{Axial couplings.}
\end{table}

\begin{figure}\label{avplot}
\resizebox{1.0\textwidth}{!}
{\includegraphics{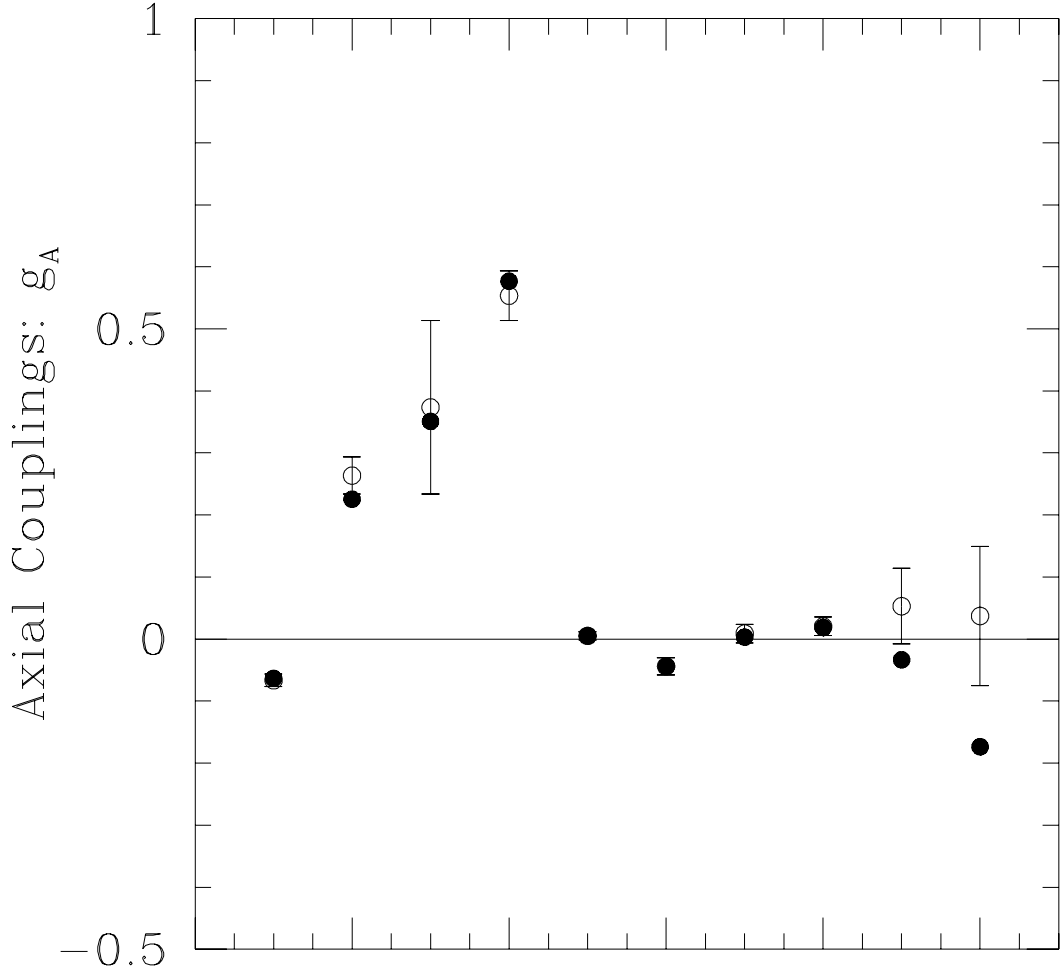}}
\caption{Deviation of the axial couplings from the best $SU(3)$-symmetric
fit. The open circles are the experimental data, and the filled circles are the
values from Fit~A discussed in Ref.~\cite{ddjm}. The points plotted are (from left to
right) $\Delta\rightarrow N$, $\Sigma^* \rightarrow \Lambda$, $\Sigma^*
\rightarrow \Sigma$, $\Xi^* \rightarrow \Xi$, $n\rightarrow p$, $\Sigma
\rightarrow \Lambda$, $\Lambda \rightarrow p$, $\Sigma \rightarrow n$, $\Xi
\rightarrow \Lambda$, and $\Xi \rightarrow \Sigma$.
}
\end{figure}

A comparison of the $1/N_c$ expansion given in Eqs.~(\ref{fitform}) and~(\ref{w})
with the experimental data
was performed in Ref.~\cite{ddjm}.  The extracted parameters from the experimental
fit are tabulated in Table~3 \footnote{The parameters used in
Eqs.~(\ref{fitform}) and~(\ref{w}) differ 
from those in Ref.~\cite{ddjm} 
because $1/N_c$ factors
have not been absorbed into the coefficients and because there is an overall factor of $2$
difference in the above formula for the axial vector currents compared to the
definition in Ref.~\cite{ddjm}.  The parameters given correspond to Fit A of
Ref.~\cite{ddjm}.}.  As discussed in Ref.~\cite{ddjm}, $c_1$ and $c_4$ are
anomalously small, and the character of the fit is not affected in any essential way
by neglecting these parameters altogether.  The coefficients $c_2$ and $c_3$ are
suppressed relative to $a_1$ and $b_2$ by a factor consistent with a power of $SU(3)$
breaking $\epsilon$.  There is evidence for the $1/N_c$ suppression factors predicted
in the $1/N_c$ expansion in the relative magnitudes of coefficients: $a_1$ and $b_2$
are comparable, as are $c_2$ and $c_3$, which is what is expected from the $1/N_c$
analysis.  However, the fit of Ref.~\cite{ddjm} is somewhat unsatisfying in that the
$\chi^2$ per d.o.f. is large, which was attributed to probable inconsistency in the
experimental data.  

A plot in Fig.~\ref{avplot} of the deviations of the baryon axial vector couplings
from an $SU(3)$-symmetric fit is revealing.   The
$SU(3)$ breaking of the baryon octet axial vector couplings obtained from hyperon
$\beta$-decay measurements is very small, as is well-known.  The $1/N_c$ expansion in
Eq.~(\ref{fitform})
predicts the axial vector couplings of the $\bf 56$ spin-flavor multiplet all together, 
which means that the $SU(3)$ breaking of hyperon 
semileptonic decay is related by spin-flavor
symmetry to the $SU(3)$ breaking of nonleptonic decay decuplet $\rightarrow$ octet +
pion.  Thus, the pion axial vector couplings between the decuplet and octet
baryons are included with the hyperon $\beta$-decay $g_A$s in the fit to $SU(3)$ breaking
of the baryon axial vector couplings.  
Fig.~\ref{avplot} reveals that
$SU(3)$ breaking is much larger for the decuplet-octet axial couplings than for
the octet-octet couplings, and that an excellent fit is obtained with the exception of the 
hyperon $\beta$ decays $\Xi \rightarrow \Lambda$, and $\Xi \rightarrow \Sigma$. 
The experimental uncertainty of these two measurements is sizeable, and there are
discrepancies between different experimental measurements for these couplings.  Thus,
it is likely that the experimental data for these decays is not entirely trustworthy,
and may account for the large $\chi^2$ per d.o.f. of the fit. Fig.~\ref{avplot} shows
that the $1/N_c$ fit favors smaller $SU(3)$ breaking for these couplings.

The first six experimentally measured baryon axial vector couplings in Fig.~8
are isovector couplings, which suggests an analysis using $SU(2) \times U(1)_Y$
flavor symmetry rather than $SU(3)$ flavor symmetry.  The $1/N_c$ expansion for
baryon isovector axial vector couplings using $SU(2)\times U(1)_Y$ flavor symmetry
is given by
\begin{eqnarray}
A^{ia} &=&  G^{ia} + {1 \over {N_c}}\left\{ N_s,  G^{ia}\right\}\nonumber\\
&&+ {1 \over {N_c}^2}\left\{ N_s, \left\{ N_s,  G^{ia}\right\}\right\}
+{1 \over {N_c}^2}\left\{ J^2,  G^{ia}\right\}
+{1 \over {N_c}^2}\left\{ I^2,  G^{ia}\right\} \nonumber\\
&&+ {1 \over {N_c}} J^i I^a +{1 \over {N_c}} J_s^i I^a
+ {1 \over {N_c}^2}\left\{ J^i, \left\{ G^{ka}, J_s^k \right\} \right\} \\
&&+ {1 \over {N_c}^2}\left\{ J^i_{s}, \left\{ G^{ka}, J_s^k \right\} \right\}
+ {1 \over {N_c}^2} \left\{ N_s, J^i I^a \right\} 
+{1 \over {N_c}^2} \left\{ N_s, J_s^i I^a \right\} \nonumber
\end{eqnarray}
where $a=1,2,3$ is an isovector index, and it is to be understood that each operator 
is multiplied by an unknown coefficient.  The matrix elements of $G^{ia}$ for baryons
with strangeness of order unity are
$O(N_c)$, whereas the matrix elements of $J^i$, $I^a$, and $J_s^i$ are $O(1)$, so a
valid truncation of the $1/N_c$ expansion is given by
\begin{equation}\label{esrdjm}
A^{ia}= a_1 G^{ia}+ a_2 {1 \over {N_c}}\left\{ N_s,  G^{ia}\right\}
\end{equation}
up to terms which are suppressed by $1/N_c^2$ relative to the leading operator
$G^{ia}$.  Eq.~(\ref{esrdjm}) yields the equal spacing rule for baryon axial
couplings derived in Ref.~\cite{djm1}.  The rule implies an equal spacing of the   
decuplet $\rightarrow$ octet baryon non-leptonic pion couplings which is linear
in strangeness,
\begin{eqnarray}
g(\Sigma^* \rightarrow \Sigma \pi) - g(\Delta \rightarrow N \pi) &=&  
g(\Xi^* \rightarrow \Xi \pi) - g(\Sigma^* \rightarrow \Sigma \pi) \\
g(\Sigma^* \rightarrow \Sigma \pi)&=&g(\Sigma^* \rightarrow \Lambda \pi) . \nonumber
\end{eqnarray}
This equal spacing rule is clearly evident in the experimental data, as shown in 
Fig.~\ref{avplot}.  Eq.~(\ref{esrdjm}) also implies $SU(4)$ spin-flavor symmetry
for the baryon isovector axial vector couplings in each strangeness sector, so 
$\beta$-decay couplings $n \rightarrow p$ and $\Sigma \rightarrow \Lambda$ are 
related to the decuplet $\rightarrow$ octet pion couplings with strangeness zero and
$-1$, respectively.  These relations are very well-satisfied.  

\subsection{Magnetic Moments}

The magnetic moment operator is $J=1$ and transforms as the $Q = T^3 + T^8/\sqrt{3}$
component of an
$SU(3)$ flavor $\bf 8$.  The $1/N_c$ expansion of the magnetic moment operator
is the same as for the axial vector couplings with
\begin{equation}
M^i = M^{i3} + \frac 1 {\sqrt{3}} \ M^{i8} \ .
\end{equation}

The isovector magnetic moments are $O(N_c)$ whereas the isoscalar magnetic moments
are $O(1)$ at leading order in the $1/N_c$ expansion, so it makes sense to construct
$1/N_c$ expansions for the isovector and isoscalar magnetic moments separately\cite{jm}.
The $1/N_c$ expansion of the isovector magnetic moments is given by
\begin{eqnarray}\label{isovector}
M^{i3} = G^{i3} + {1 \over N_c} \left\{N_s, G^{i3} \right\},
\end{eqnarray}
up to terms  which are suppressed by $1/N_c^2$ relative to 
leading $1$-body operator $G^{i3}$.  The $1/N_c$ expansion of the isoscalar magnetic moments is given by
\begin{eqnarray}\label{isoscalar}
M^{i8} = J^i + J_s^i + {1 \over N_c} \left\{N_s, J^i \right\}
+ {1 \over N_c} \left\{N_s, J^i_s \right\},
\end{eqnarray}
up to terms of order $1/N_c^2$ compared to the two leading order $1$-body operators
$J^i$ and $J_s^i$.  It is to be understood that every operator in
Eqs.~(\ref{isovector})
and~(\ref{isoscalar}) is multiplied by an unknown coefficient of order unity.  

\begin{table}\label{ivis}
\begin{tabular}{ccccc}
\hline
\tablehead{1}{c}{b}{  }
&\tablehead{1}{c}{b}{Isovector}
&\tablehead{1}{c}{b}{$1/N_c$}
&\tablehead{1}{c}{b}{Flavor}
&\tablehead{1}{c}{b}{Expt.} \\
\hline
\smallskip
V1 & $(p-n)-3(\Xi^0-\Xi^-)=2(\Sigma^{+} - \Sigma^-)$ &$1/N_c$ &$\surd$ &$10\pm2\%$ \\
V2 & $\Delta^{++}-\Delta^-=\frac95(p-n)$&$1/N_c$ &$\surd$ & \\
V3 & $\Lambda\Sigma^{*0}=- \sqrt2 \Lambda \Sigma^{0}$&$1/N_c$ &$\surd$ & \\
V4 & $\Sigma^{*+}-\Sigma^{*-} = \frac32 (\Sigma^{+} - \Sigma^-)$&$1/N_c$ &$\surd$ & \\
V5 & $\Xi^{*0}-\Xi^{*-} = -3(\Xi^0-\Xi^-)$&$1/N_c$ &$\surd$ & \\
V6 & $\sqrt2(\Sigma\Sigma^{*+}-\Sigma\Sigma^{*-}) =(\Sigma^{+} - \Sigma^-)$ 
&$1/N_c$ &$\surd$ & \\
V7 & $\Xi\Xi^{*0}-\Xi\Xi^{*-}=-2\sqrt2(\Xi^0-\Xi^-)$
&$1/N_c$ &$\surd$ & \\
\hline
V8 & $-2\Lambda\Sigma^0=(\Sigma^+-\Sigma^-)$ &$1/N_c$ &$\surd$ & $11\pm5\%$ \\
V9 & $p\Delta^++n\Delta^0=\sqrt2(p-n)$ &$1/N_c$ &$\surd$ & $3\pm3\%$ \\
\hline
V10${}_1$ &$(\Sigma^+-\Sigma^-)=(p-n)$ &$1$ &$\surd$ & $27\pm1\%$ \\
V10${}_2$ &$(\Sigma^+-\Sigma^-)=\left(1-{1\over N_c}\right)(p-n)$
&$1$ &$\epsilon$ & $13\pm2\%$ \\
\hline
\tablehead{1}{c}{b}{  }
&\tablehead{1}{c}{b}{Isoscalar}
&\tablehead{1}{c}{b}{}
&\tablehead{1}{c}{b}{}
&\tablehead{1}{c}{b}{} \\
\hline
\smallskip
S1 & $(p+n)-3(\Xi^0+\Xi^-) = - 3 \Lambda + \frac32(\Sigma^{+} +
\Sigma^-)-\frac43\Omega^-$ & $1/N_c^2$ &$\surd$ &$4\pm5\%$ \\
S2 & $\Delta^{++}+\Delta^-=3(p+n)$& $1/N_c^2$ &$\surd$ & \\
S3 & $\frac23 (\Xi^{*0}+\Xi^{*-})=\Lambda+\frac32(\Sigma^{+} + \Sigma^-)
-(p+n)+(\Xi^0+\Xi^-)$& $1/N_c^2$ &$\surd$ & \\
S4 & $\Sigma^{*+}+ \Sigma^{*-} =
\frac32(\Sigma^{+} + \Sigma^-)+3\Lambda$ & $1/N_c^2$ &$\surd$ & \\
S5 & $\frac{3}{\sqrt2}(\Sigma\Sigma^{*+}+\Sigma\Sigma^{*-}) =
3(\Sigma^{+} +\Sigma^-)- (\Sigma^{*+} +\Sigma^{*-})$ & $1/N_c^2$ &$\surd$ & \\
S6 & $\frac{3}{\sqrt2}(\Xi\Xi^{*0}+\Xi\Xi^{*-})=-3(\Xi^0+\Xi^-)
+(\Xi^{*0}+\Xi^{*-})$ & $1/N_c^2$ &$\surd$ & \\
S7 & $5(p+n) - (\Xi^0+\Xi^-)= 4(\Sigma^+ + \Sigma^-)$ & $1/N_c$ &$\surd$ & $22 \pm 4
\%$ \\
S8 & $(p+n) - 3 \Lambda = \frac 1 2(\Sigma^+ + \Sigma^-)- (\Xi^0+\Xi^-)$ & $1/N_c$
&$\epsilon$ & $7 \pm 1 \%$ \\
\hline
\tablehead{1}{c}{b}{  }
&\tablehead{1}{c}{b}{Isoscalar/Isovector Relations}
&\tablehead{1}{c}{b}{}
&\tablehead{1}{c}{b}{}
&\tablehead{1}{c}{b}{} \\
\hline
\smallskip
S/V${}_1$ & $(\Sigma^++\Sigma^-)-\frac12(\Xi^0+\Xi^-) = \frac12(p+n)+
3\left({1\over N_c}-{2\over N_c^2}\right)(p-n)$ &$1$ &$\epsilon$ &$10\pm3\%$\\
&$\Delta^{++}=\frac32(p+n)+\frac{9}{10}(p-n)$&$1/N_c^2$&$\surd$& $21\pm10\%$\\
\hline
\end{tabular}
\caption{Baryon Magnetic Moments in the $1/N_c$ expansion. 
The isovector magnetic moments are 
$O(N_c)$ at leading order, and the isoscalar magnetic moments are $O(1)$. 
A $\surd$ implies that the relation is satisfied to that order in $1/N_c$ 
{\it to all orders in $SU(3)$ breaking}. The experimental 
accuracies are given for the relations whose magnetic moments 
have been measured.
}
\end{table}

There are 21 independent magnetic moments of the baryon octet and
decuplet, including transition magnetic moments.  These 21 magnetic moments consist
of 11 isovector combinations and 10 isoscalar combinations.  The $1/N_c$ hierarchy
of combinations of isovector and isoscalar magnetic moments is given in
Table~\ref{ivis}.  

The 11 isovector magnetic moment combinations are parametrized in
terms of the two operators of Eq.~(\ref{isovector}), so there are nine isovector
magnetic moment relations which are satisfied to order $1/N_c$.  These relations are
listed as $V1-9$ in Table~\ref{ivis}.  Only one combination is measured, and the
experimental accuracy $10 \pm 2 \%$ is consistent with the $1/N_c^2$ prediction of
the $1/N_c$ expansion.  The $1/N_c$ expansion of the isovector magnetic moments can
be truncated to the single operator $G^{ia}$ by eliminating the subleading operator
$\left\{N_s, G^{i3} \right\}$ operator.  The isovector magnetic moment combination
corresponding to this subleading operator is $O(1)$ in the $1/N_c$ expansion, or of
relative order $1/N_c$ compared to the leading $O(N_c)$ contribution, and is listed
as ${V10}_1$ is Table~\ref{ivis}.  The experimental accuracy of this relation is
$27 \pm 1 \%$, which is consistent with the prediction $1/N_c$ of the $1/N_c$
hierarchy.  It is possible to derive a slightly different version of this mass
combination by considering an $SU(3)$ analysis.  In this analysis, the $2$-body
operator is $\left\{T^8, G^{i3} \right\}$, which is first order in $SU(3)$ breaking
and order $1/N_c$ compared to the leading operator.  
The magnetic moment combination corresponding to this $SU(3)$ operator is listed as
${V10}_2$.  The experimental accuracy of this relation is $13 \pm 2 \%$, which is
completely consistent with the theoretic prediction of $\epsilon/N_c$ of the $1/N_c$
expansion.

The 10 isoscalar magnetic moment combinations are parametrized by two $1$-body
operators at leading order in the $1/N_c$ expansion, so there are eight isoscalar
magnetic moment relations, which appear as $S1-8$ in Table~\ref{ivis}.  The $1/N_c$
expansion of Eq.~(\ref{isoscalar}) contains four operators, so there are six
isoscalar combinations $S1-6$ which are order $1/N_c^2$.  The two subleading
$2$-body operators correspond to isoscalar relations $S7$ and $S8$, which are order
$1/N_c$.  In addition, $S8$ is first order in $SU(3)$ breaking.  The experimental
accuracies of the isoscalar magnetic moment combinations are in complete accord with
the $1/N_c$ hierarchy.

Finally, there are two additional relations given in Table~\ref{ivis}.
$S/V_1$ is a relation normalizing the isovector magnetic 
moments to the
isoscalar magnetic moments in the $SU(3)$ flavor symmetry limit, and the 
last relation predicting the
$\Delta^{++}$ magnetic moment is a linear combination of $V2$ and $S2$.

\begin{table}
\caption{Magnetic moments. \label{MagP}}
\begin{tabular}{cc}
\hline
& Fit A \\ 
\hline
$a_1$          & 5.614 $\pm $ 0.122 \\
$b_2$          & 0.216 $\pm $ 0.354 \\
$d_3$          &3.753 $\pm $ 0.639 \\
$c_1$        &-1.092 $\pm $ 0.230 \\
$c_2$        & 0.612 $\pm $ 0.276 \\
$\delta c_2$ & 0.066 $\pm $ 0.258 \\
$c_3$        &-0.522 $\pm $ 0.222 \\
$\delta c_3$ & 0.024 $\pm $ 0.312 \\
$c_4$        & 0.258 $\pm $ 0.228 \\
$\delta c_4$ &-0.288 $\pm $ 0.180 \\
\end{tabular}
\end{table}

\begin{figure}\label{magmom}
\resizebox{1.0\textwidth}{!}
{\includegraphics{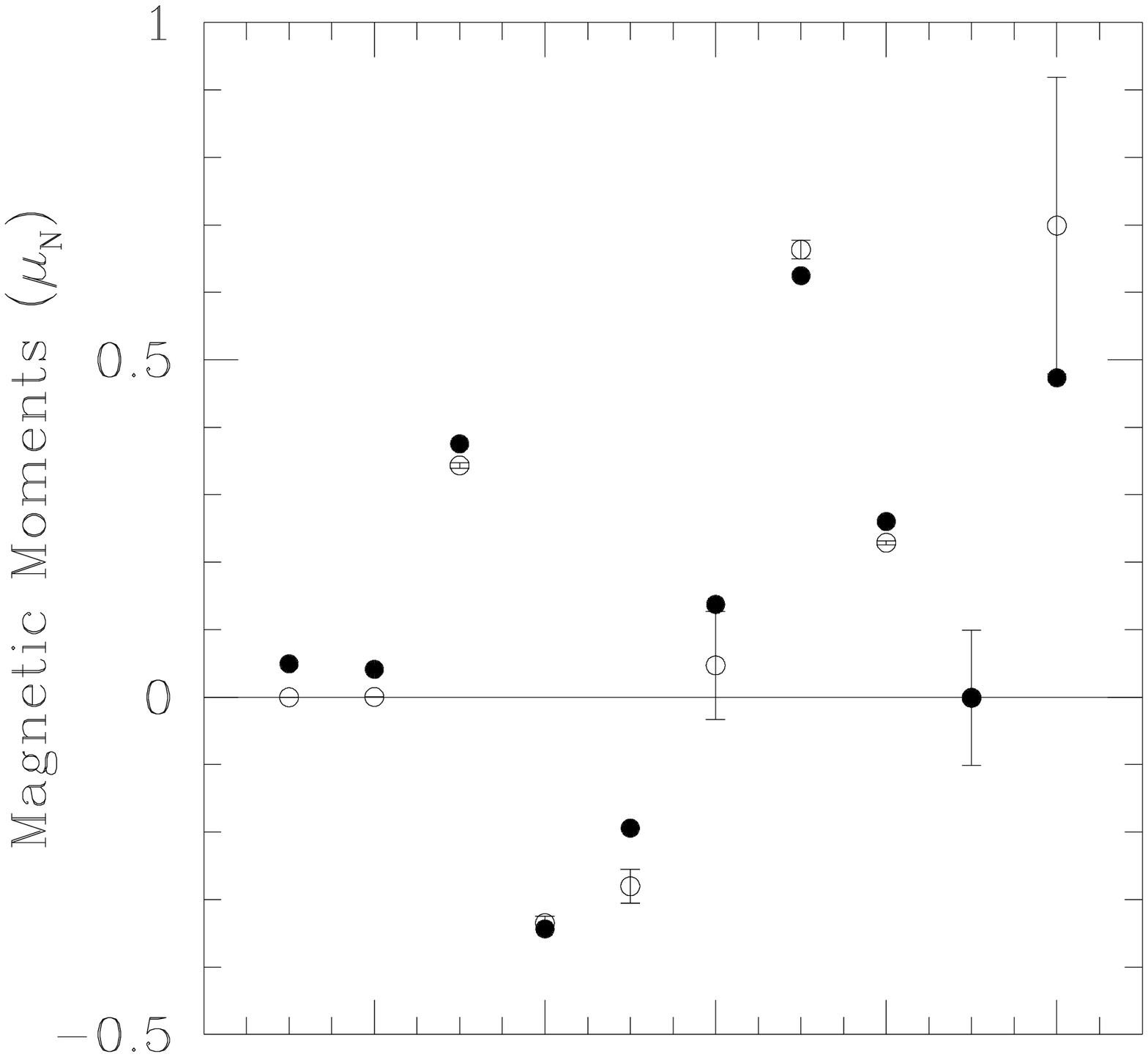}}
\caption{Deviation of the magnetic moments from the best $SU(3)$-symmetric
fit. The open circles are the experimental data, and the filled circles are the
values from Fit~A discussed in the text. The order of the magnetic moments is
$p$, $n$, $\Lambda$, $\Sigma^+$, $\Sigma^-$, $\Sigma^0 \Lambda$, $\Xi^0$, $\Xi^-$,
$p \Delta^+$, and $\Omega$. The $\Delta^{++}$ magnetic moment
has not been plotted, since the experimental value has a very large error.
}
\end{figure}

\begin{figure}\label{mmloop}
\resizebox{1.0\textwidth}{!}
{\includegraphics{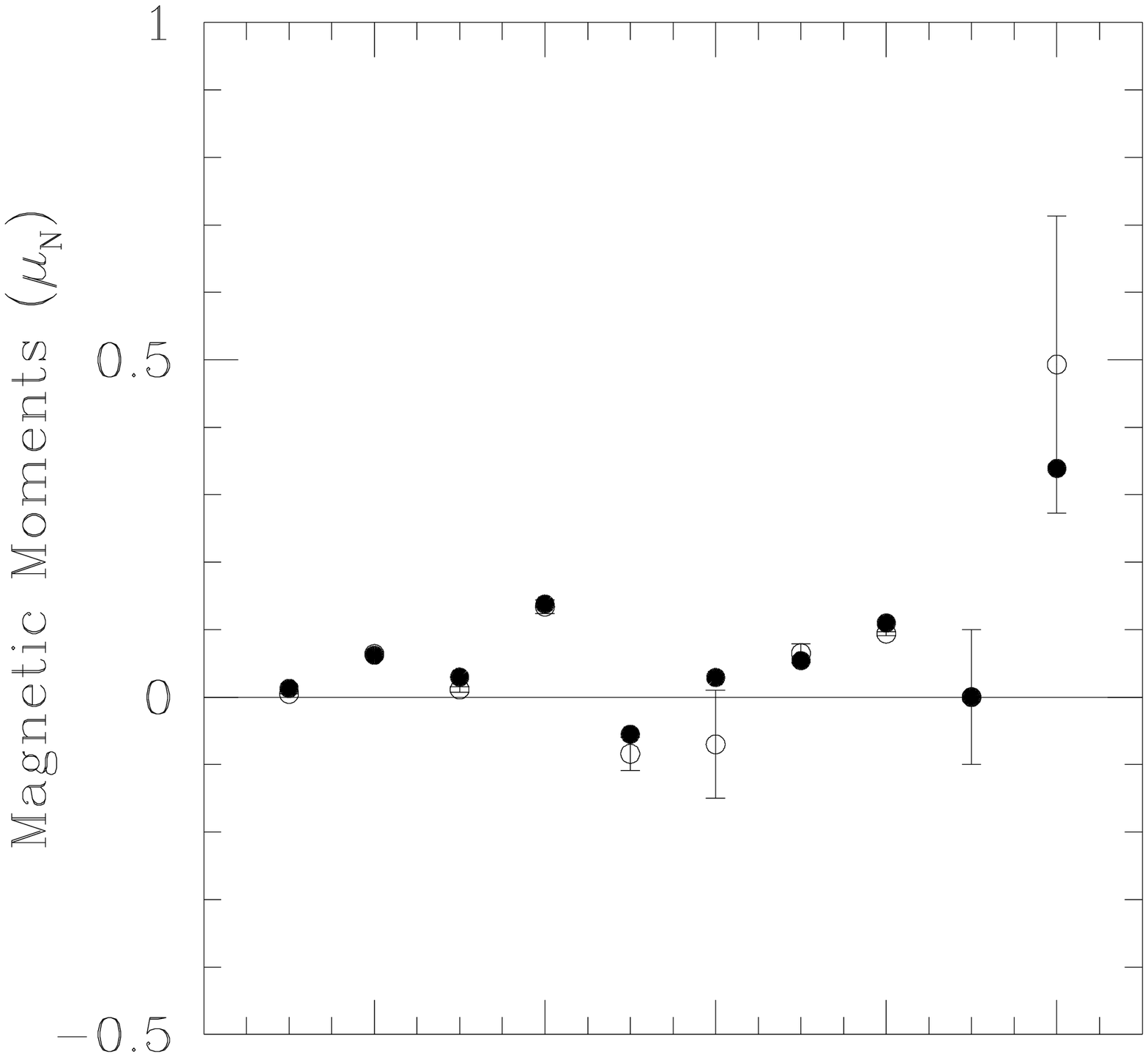}}
\caption{Deviation of the magnetic moments from the best $SU(3)$-symmetric
fit plus the leading $O(\sqrt{m_s})$ chiral loop correction. The deviations
should be compared with those in Fig.~\ref{magmom}.
}
\end{figure}

An alternative approach to the magnetic moments is possible using the $1/N_c$
expansion with $SU(3)$ flavor symmetry breaking.  
The same formula derived for the
baryon axial vector currents applies since the magnetic moments also transform as
$J=1$ and as a component of an $SU(3)$ octet in the flavor symmetry limit.  The
analysis of flavor symmetry breaking involves the same representations analyzed for
the baryon axial vector couplings, so Eq.~(\ref{fitform}) can be applied to the
magnetic moments.  
A fit to the baryon magnetic moments using Eq.~(\ref{fitform}) gives 
the parameters listed in Table~\ref{MagP} taken from 
Ref.~\cite{ddjm}\footnote{Again, the parameters used in
Eqs.~(\ref{fitform}) and~(\ref{w}) differ 
from those in Ref.~\cite{ddjm} 
because $1/N_c$ factors
have not been absorbed into the coefficients and because there is an overall factor of $2$
difference in the above formula for the axial vector currents compared to the
definition in Ref.~\cite{ddjm}.  The parameters given correspond to Fit A of
Ref.~\cite{ddjm}.}.
For the magnetic moments, the extracted value of $b_2$ is small, which implies that
$F/D$ is very close to the $SU(6)$ symmetry prediction of $2/3$.  
A plot of the deviations of the baryon magnetic moments 
from an $SU(3)$-symmetric fit, given in Fig.~\ref{magmom}, shows that $SU(3)$
breaking is considerably larger for the magnetic moments than for the baryon axial
vector couplings.  Furthermore, $SU(3)$ symmetry breaking for the magnetic moments is
dominated by an $O(N_c \sqrt{{m_s}})$ chiral loop correction, as shown in 
Fig~\ref{mmloop} where the deviation of the baryon magnetic moments from an 
$SU(3)$-symmetric fit together with the leading chiral loop correction is plotted.  
Clearly, the remaining $SU(3)$ breaking in the magnetic moments is much reduced when
the leading non-analytic correction is included in the fit.  This result also can be
seen in Table~~\ref{MagP} in terms of the small values of the extracted parameters
$\delta c_{2-4}$ which measure the deviation of the extracted $SU(3)$ breaking
parameters $c_{2-4}$ 
from the flavor symmetry breaking structure given by the dominant 
chiral loop graph.  There is no analogue of
this chiral non-analytic correction for the baryon axial vector currents, so
the $SU(3)$ breaking patterns of the baryon magnetic
moments and the baryon axial vector currents are not similar.

\section{Conclusions}

The $1/N_c$ expansion for QCD baryons is both useful and predictive.  In the formal
large-$N_c$ limit, there is a spin-flavor symmetry for baryons.  For finite $N_c$,
the spin and flavor structure of the baryon $1/N_c$ expansion is prescribed at each
order in $1/N_c$.  The $1/N_c$ expansion is given in terms of operator
products of the generators of the baryon spin-flavor algebra which transform in a
certain manner under spin $\otimes$ flavor symmetry.  The order in $1/N_c$ of each
operator structure is determined in the $1/N_c$ expansion, so the $1/N_c$ expansion
predicts a hierarchy of spin and flavor relations for baryons in $1/N_c$.  
The predicted hierarchy of the $1/N_c$ expansion is evident in the
baryon masses, axial vector currents and magnetic moments.  The pattern
of spin-flavor symmetry breaking is quite intricate since $1/N_c$ and 
$SU(3)$ flavor symmetry breaking are comparable in QCD.   The presence of $1/N_c$
suppression factors explains why $SU(3)$ flavor symmetry works to a greater accuracy
for baryons than predicted from an analysis of $SU(3)$ breaking alone, and gives
a quantitative
understanding of spin-flavor symmetry breaking for QCD baryons.

\begin{theacknowledgments}
I wish to thank J\"urgen Engelfried and Mariana Kirchbach 
for organizing such a wonderful workshop.  Special thanks to 
Ruben Flores-Mendieta for his hospitality, and to many of the participants for
interesting discussions and experiences.  This work was supported in part
by the U.S. Department of Energy, under Grant No. DOE-FG03-97ER40546. 
\end{theacknowledgments}

\end{document}